\documentstyle[twocolumn,prl,aps]{revtex}
\begin{document}
\preprint{OKHEP-96-05}
\draft
\title{\hfill OKHEP-96-05\\
Comment on ``Sonoluminescence as Quantum Vacuum Radiation''}
\author{Kimball A. Milton\cite{byline}}
\address{Department of Physics and Astronomy, The University of
Oklahoma,
Norman, OK 73019-0225}
\date{\today}
\maketitle
\pacs{78.60.Mq, 42.50.Lc, 12.20.Ds, 03.70.+k}

In a recent Letter \cite{eberlein} Eberlein proposed that the
dynamical
Casimir effect \cite{js} or perhaps more properly the Unruh effect
\cite{unruh}
could be responsible for the remarkable light emission during
single-bubble
sonoluminescence \cite{sono}.  However, I believe that the effect
proposed
is far too small to account for the observation of something like a
million
optical photons per bubble collapse.  This is supported by detailed
calculations
which will appear elsewhere \cite{milton}. Moreover, even the model
profile
given in \cite{eberlein}, although implying supraluminal velocities,
yields
an energy output orders of magnitude too small.  Although technical
objections to \cite{eberlein} are given in \cite{milton},
having to do with the force on a dielectric surface, I will
concentrate
here on the insufficiency of the energy radiated in such a model.

One would expect that the relevant time scales for a macroscopic,
collapsing
bubble to be much longer than the period of visible light,
$t\sim10^{-15}$ s.
Indeed, the observed collapse time for sonoluminescing bubbles is
$\sim10^{-6}$ s,
while the duration of the flash is $\lesssim 10^{-11}$ s.  If that is
the
case, then an adiabatic approximation should be highly accurate.
Statically, the quantum Casimir energy of electromagnetic field
fluctuations in a bubble of radius $a\sim 10^{-4}$--$10^{-3}$ cm in a
liquid
should be of order
\begin{equation}
E_c\sim\frac{\hbar c}{a}\sim 10^{-1} \mbox{eV},
\end{equation}
some 8 orders of magnitude too small to be relevant.  In fact,
putting in the numbers \cite{milton} reduces the energy by another
3 orders of magnitude.  It should be emphasized that formally
divergent results which are therefore highly sensitive to
cutoffs \cite{js,chodos} are not physically plausible.

A reliable estimate for the power radiated should be obtainable from
the
Larmor formula,
\begin{equation}
P=\frac{2}{3}\frac{(\ddot d)^2}{c^3},
\end{equation}
where $d$ is the dipole moment.  A reasonable estimate for the latter
is $d\sim ea$ (in fact, the short-wavelength limit given in
\cite{eberlein}
 is equivalent to $d\approx e a\dot a/c$), so we would expect that
the energy emitted during a flash of duration $\tau$ to be roughly
\begin{equation}
E\sim \alpha\hbar c\frac{a^2}{c^3\tau^3}\sim 10^{-44} \mbox{eV}
 \mbox{s}^3/\tau^3.
\end{equation}
Even for $\tau$ as short as a femtosecond, only 10 eV of energy is
radiated.
An extraordinarily short time scale, $\tau\sim 10^{-17}$ s, is
required to
liberate $10^7$ eV.

One would think such a time scale would imply supraluminal
velocities,
$a/\tau\sim10^{13}$ cm/s.  Indeed, the specific model proposed by
\cite{eberlein}
has precisely this feature, and even so yields only $10^3$ eV of
energy.
But, it is possible to imagine that velocities could remain
nonrelativistic
while the acceleration, or the derivative thereof, becomes very
large.
Precisely such phenomena occur during the formation of a shock.
Classical shock models of sonoluminescence have been proposed
\cite{green}.
In this case, the radiation is supposed to be emitted by
bremsstrahlung
after ionization of the air in the bubble.  Whether or not such a
picture
is viable, it is apparent that it has nothing to do with quantum
radiation.

This work was supported in part by the US Department of Energy.
I thank C. Bender, C. Eberlein, J. Ng, and D. Sciama for useful
conversations.

\end{document}